\documentclass[10pt,journal,compsoc]{IEEEtran}

\usepackage[colorlinks,urlcolor=blue,linkcolor=blue,citecolor=blue]{hyperref}

\usepackage{mathtools}
\usepackage{booktabs}
\usepackage{url}
\usepackage{amssymb}
\usepackage{listings}
\usepackage{tikz}

\graphicspath{{./figures/}}

\definecolor{mygreen}{rgb}{0,0.6,0}
\definecolor{mygray}{rgb}{0.5,0.5,0.5}
\definecolor{mymauve}{rgb}{0.58,0,0.82}
\definecolor{altblue}{rgb}{0.0,0.6,1.0}
\definecolor{lstbg}{gray}{0.9}
\lstdefinelanguage[firedrake]{python}[]{python}{%
  keywordstyle={[2]\color{red}},
   morekeywords={[2]UnitCubeMesh,MeshHierarchy,FunctionSpace,Function,TrialFunction,TestFunction,DirichletBC,SpatialCoordinate,Constant,solve},
  keywordstyle={[3]\color{orange!50!black}},
  morekeywords={[3]grad,dx,inner,pi,sin,cos,tan}
}
\lstdefinelanguage[highlighting]{python}[firedrake]{python}{
	moredelim=**[is][{\btHL[fill=red!30,draw=black,thin]}]{`mesh*}{`},
	moredelim=**[is][{\btHL[fill=orange!30,draw=black,thin]}]{`fs*}{`},
	moredelim=**[is][{\btHL[fill=mygreen!30,draw=black,thin]}]{`bcs*}{`},
	moredelim=**[is][{\btHL[fill=blue!30,draw=black,thin]}]{`rhs*}{`},
	moredelim=**[is][{\btHL[fill=violet!30,draw=black,thin]}]{`bilf*}{`}
}

\tikzset{term/.style={draw=none,fill=none,rectangle,inner sep=0pt,anchor=base}}
\tikzstyle{every picture}+=[remember picture]
\everymath{\displaystyle}

\makeatletter

\newcommand\mathterm[2][]{%
  \@ifnextchar[{\@mathtermopts{#1}{#2}}{\@mathtermnoopts{#1}{#2}}}
\def\@mathtermnoopts#1#2{%
  \tikz [baseline] { \node [term] (#1) {$#2$}; }}
\def\@mathtermopts#1#2[#3]{%
  \tikz [baseline] { \node [rectangle,inner sep=2pt,rounded corners=2pt,anchor=base,,#3] (#1) {$#2$}; }}

\newcommand\textterm[2][]{%
  \@ifnextchar[{\@texttermopts{#1}{#2}}{\@texttermnoopts{#1}{#2}}}
\def\@texttermnoopts#1#2{%
  \tikz [baseline] { \node [term] (#1) {#2}; }}
\def\@texttermopts#1#2[#3]{%
  \tikz [baseline] { \node [rectangle,inner sep=2pt,rounded corners=2pt,anchor=base,,#3] (#1) {#2}; }}

\newcommand\indicate[2][black]{%
  \tikz [baseline] \node [inner sep=0pt,anchor=base] (i#2) {\vphantom|};
  \@ifnextchar[{\@indicateopts{#1}{#2}}{\@indicatenoopts{#1}{#2}}}
\def\@indicatenoopts#1#2{%
  {\color{#1} \tikz[overlay] \path[line width=1pt,draw=#1,-stealth] (i#2) edge (#2);}}
\def\@indicateopts#1#2[#3]{%
  {\color{#1} \tikz[overlay] \path[line width=1pt,draw=#1,-stealth] (i#2) [#3] edge (#2);}}

\newenvironment{btHighlight}[1][]
{\begingroup\tikzset{bt@Highlight@par/.style={#1}}\begin{lrbox}{\@tempboxa}}
{\end{lrbox}\bt@HL@box[bt@Highlight@par]{\@tempboxa}\endgroup}

\newcommand\btHL[1][]{%
  \begin{btHighlight}[#1]\bgroup\aftergroup\bt@HL@endenv%
}
\def\bt@HL@endenv{%
  \end{btHighlight}%   
  \egroup
}
\newcommand{\bt@HL@box}[2][]{%
  \tikz[#1]{%
    \pgfpathrectangle{\pgfpoint{1pt}{0pt}}{\pgfpoint{\wd #2}{\ht #2}}%
    \pgfusepath{use as bounding box}%
    \node[anchor=base west, fill=orange!30,outer sep=0pt,inner xsep=1pt, inner ysep=0pt, rounded corners=2pt, minimum height=\ht\strutbox+1pt,#1]{\raisebox{1pt}{\strut}\strut\usebox{#2}};
  }%
}
\makeatother

\begin{document}
\urlstyle{}
%\sptitle{Department: Head}
%\editor{Editor: Name, xxxx@email}

\title{Code generation for productive portable scalable finite element simulation in Firedrake.}

\author{Jack D.~Betteridge, Patrick E.~Farrell and David A.~Ham
\IEEEcompsocitemizethanks{\IEEEcompsocthanksitem J.~D.~Betteridge and D.~A.~Ham are at Imperial College London. \IEEEcompsocthanksitem P.~E.~Farrell is at the University of Oxford.}}

%\markboth{Department Head}{Paper title}

\IEEEtitleabstractindextext{\begin{abstract}
Creating scalable, high performance PDE-based simulations requires a suitable combination of discretizations, differential operators, preconditioners and solvers. The required combination changes with the application and with the available hardware, yet software development time is a severely limited resource for most scientists and engineers. Here we demonstrate that generating simulation code from a high-level Python interface provides an effective mechanism for creating high performance simulations from very few lines of user code. We demonstrate that moving from one supercomputer to another can require significant algorithmic changes to achieve scalable performance, but that the code generation approach enables these algorithmic changes to be achieved with minimal development effort.
\end{abstract}}
 
\maketitle

\IEEEraisesectionheading{\section{Introduction}\label{sec:introduction}}

\IEEEPARstart{T}{he simulation} of continuum mechanics problems through the numerical solution of partial differential equations (PDEs) is a classical high performance computing task. The three-dimensional multiscale nature of science and engineering challenges in fields such as fluid mechanics, solid mechanics, and electromagnetics means that the number of variables and operations required can become very large very quickly. The ubiquitous nature of these systems means that many scientists and engineers need to solve problems in this class. However, the specific equations to be solved, the discretizations and solver strategies employed, and the coupling of these to other processes, all vary greatly from case to case. It is also seldom the case that solving a PDE is the whole problem: scientists and engineers need to quantify uncertainty in solutions, compute their sensitivity to inputs, assimilate observed data, or optimize parameters.

The combination of all of these factors means that bespoke simulations are the norm: most simulations are put together for a particular purpose by one or a small group of scientists or engineers, and run a limited number of times before the problem specification changes in a way which demands alterations to the equations being solved or the algorithms being used to solve them. To add to this complexity, these users will rarely be able to design or procure their own supercomputer. Rather, they will need to make effective use of whichever institutional or national facilities they can access, making the required algorithmic changes as they move from one machine to another. Even the emergence of cloud computing resources offers little comfort, since the ever-changing hardware landscape means that the performance characteristics of processors such as memory bandwidth, core count, and vector length evolve rapidly and may vary between providers, or even between the classes of cloud node suitable for simulations of different sizes.

Making high performance parallel simulation viable for this long tail of simulation science and engineering is a different proposition from the large-scale operational simulations carried out in fields such as weather forecasting and seismic inversion, in which the same computational problem is solved an enormous number of times for different data inputs. In those operational fields, the cost of laborious implementation and optimization work can be amortized over countless simulation runs so that the cost per simulation is not dominated by development. Conversely, for the typical simulation scientist, the requirement is for high performance for as close to zero programmer effort as can be achieved. Further, the cost of changing any aspect of the simulation specification or algorithm must be minimal, lest the code become fossilized and unable to adapt to the next change in task or available hardware.

Here we demonstrate the utility of the Firedrake automated finite element system in addressing this need. Firedrake enables users to write high-level Python code in a symbolic language which reflects the mathematics of the problem. It integrates tightly with the Portable Extensible Toolkit for Scientific Computation (PETSc) \cite{petsc-user-ref} to enable runtime programmable preconditioners and solvers that reach back to the discretization as required. We will demonstrate the capabilities of Firedrake in providing productive performance portability by taking a fluids simulation tuned for ARCHER, the previous UK national supercomputer, and porting it to the ARM-based Isambard and ARCHER2, the AMD-based new UK national supercomputer. We find that algorithmic choices made for one platform are suboptimal on other platforms, but that working from a high-level mathematical specification of the problem and generating the implementation automatically enables the necessary algorithmic changes to be made at minimal user cost. 

%%%%%%%%%%%%%%%%%%%%%%%%%%%%%%
\section{Firedrake}
%%%%%%%%%%%%%%%%%%%%%%%%%%%%%%

The Firedrake automated finite element system \cite{rathgeber2016firedrake} (\url{https://firedrakeproject.org}) is a Python package which generates numerical solutions to PDEs from a very high level mathematical specification provided by the user. The PDEs to be solved by Firedrake are specified in the Unified Form Language (UFL) \cite{alnaes2014}, a specialized computer algebra language also employed by FEniCS \cite{logg2012} and DUNE \cite{bastian2021}. Firedrake is distinguished from those projects by its pure Python implementation, with a greater emphasis on code generation to deliver high performance, and by its tight integration with the linear and nonlinear solver capabilities of PETSc.

The operation of Firedrake is best illustrated with an example. We consider the Poisson equation in three dimensions, which is simple to write and which also acts as a model for the fluid dynamics problem which will be our target application. The direct correspondence between the mathematics of the problem and the user code is illustrated by highlighting the terms which which correspond to lines of code in Listing \ref{lst:poisson}, which is a minimal functional Firedrake script which solves the Poisson equation in 3D. 

We consider Poisson's equation in a \textterm[mesh]{unit cube domain}[fill=red!30,draw=black,thin], $\Omega=[0,1]^3\subset\mathbb{R}^3$ on a $32\times 32\times 32$ tetrahedral mesh which is further refined twice to create a hierarchy of 3 nested meshes which we will use in a multigrid solver.
We use a \textterm[fs]{degree 3 continuous (CG) finite element space }[fill=orange!30,draw=black,thin]
We solve the Poisson equation subhect to homogeneous \textterm[bcs]{(zero) Dirichlet boundary conditions (BCs):}[fill=mygreen!30,draw=black,thin]\\
\begin{equation}
\left\{
\begin{aligned}
	-\nabla^2 u &= f && \text{on } \Omega,\\
	u &= 0 && \text{on } \partial\Omega,
\end{aligned}
\right.
\label{eq:poisson}
\end{equation}
where the right hand side $f$ is given by:
\begin{equation}
\mathterm[rhs]{
\begin{array}{l}
	f\!\left(
\begin{smallmatrix}
	\tiny
	x\\ y\\ z
\end{smallmatrix}\right) = -\frac{\pi^2}{2}\\
	\hspace{3.5em}\times\left( 2\cos(\pi x) - \cos\left( \frac{\pi x}{2} \right)\right. \\
	\hspace{4.5em} \left. - 2(a^2 + b^2)\sin(\pi x)\tan \left( \frac{\pi x}{4} \right)  \right)\\
	\hspace{3.5em}\times\sin(a\pi y) \sin(b\pi z).
\end{array}}[fill=blue!30,draw=black,thin]
\end{equation}
This has analytic solution
\begin{equation}
\begin{array}{l}
	u\!\left(
\begin{smallmatrix}
	\tiny
	x\\ y\\ z
\end{smallmatrix}\right) = \sin(\pi x)\tan\left(\frac{\pi x}{4}\right)\\
	\hspace{4.5em}\times\sin(a\pi y)\sin(b\pi z),
\end{array}
\end{equation}
which we use to calculate the error in our numerical solution.

Multiplying by a test function $v$, from the same finite element space, and integrating by parts yields the prototypical weak form: find $u\in V$ such that
\begin{equation}
	\hspace{4em}\mathterm[bilf]{\int_\Omega \nabla u\cdot \nabla v\ dx = \int_\Omega f v\ dx}[fill=violet!30,draw=black,thin] \qquad \forall v \in V.
\end{equation}

The \lstinline+solve+ call at line \ref{line:solve} creates a PETSc solver object which solves the discretized system (a linear system, in this case). PETSc needs callback functions which assemble the left hand side matrix and residual of the weak form. Firedrake's compiler creates optimized C representations of these operations from the UFL provided, and passes the compiled C callbacks to PETSc. The PETSc solver is fully programmable using the supplied dictionary of parameters, and this can include further delegation of preconditioner stages back to Firedrake. For example, line \ref{line:pcpatch} of Listing \ref{lst:fmg} delegates the preconditioning of the multigrid smoother back to the PatchPC preconditioner presented in \cite{farrell2019pcpatch}. Optimal preconditioners for PDE problems usually depend on discretization-specific information. The tight coupling of Firedrake and PETSc enables this mathematical fact to be easily translated into executable code: PETSc controls the solving process and calls back to Firedrake whenever discretization-specific code is required. The Firedrake preconditioners can themselves contain (preconditioned) linear solves for subsystems and will call back to PETSc to execute these.

A notable absence from listing \ref{lst:poisson} is any explicit parallel code: there are no calls to MPI or threading directives. In fact, none are needed. The script presented will run in parallel without modification, simply by running it under MPI. The user code specifies the mathematical problem to be solved, while Firedrake decides how to execute it on the parallel architecture.
%The mesh and associated function space and field data will be automatically decomposed and the solver will execute in parallel.  

\begin{lstlisting}[language={[highlighting]python}, label={lst:poisson}, caption={Simple Firedrake script for solving the Poisson equation in 3D, Listing \ref{lst:fmg} and Listing \ref{lst:telescope} give example solver parameters that can be used in line \ref{line:solverparam}.}, float={t}]
from firedrake import *

N = 32 #*\label{line:meshresolution}*
`mesh*mesh = UnitCubeMesh(N, N, N)`
hierarchy = MeshHierarchy(mesh, 2) #*\label{line:meshhierarchy}*
mesh = hierarchy[-1]

`fs*degree = 3` #*\label{line:degree}*
`fs*V = FunctionSpace(mesh, "CG", degree)`
u = TrialFunction(V)
v = TestFunction(V)

`bcs*bcs = DirichletBC(V, zero(),`
  `bcs*(1, 2, 3, 4, 5, 6))`

x, y, z = SpatialCoordinate(mesh)
a = Constant(1)
b = Constant(2)

`rhs*f = -pi**2 / 2`
`rhs*f *= 2*cos(pi*x) - cos(pi*x/2)`
  `rhs* - 2*(a**2 + b**2)*sin(pi*x)*tan(pi*x/4)`
`rhs*f *= sin(a*pi*y)*sin(b*pi*z)`

`bilf*a = dot(grad(u), grad(v))*dx`
`bilf*L = f*v*dx`

solver_parameters = {...} #*\label{line:solverparam}*
u = Function(V)
solve(`bilf*a == L`, u, bcs=bcs, solver_parameters=solver_parameters) #*\label{line:solve}*
\end{lstlisting}

\lstset{language=python,moredelim=[is][\color{red}]{--}{--},moredelim=[is][\color{mygreen}]{++}{++}}
\begin{lstlisting}[float={t}, caption={PETSc solver options for the Poisson problem. These options use a full multigrid method smoothed with 2 Chebyshev iterations at each level. The preconditioner of the smoother is delegated back to Firedrake's patch preconditioner, and the coarse problem is solved with a direct (LU) solver.}, label={lst:fmg}]
solver_parameters = {
  "ksp_type": "preonly",
  "pc_type": "mg",
  "pc_mg_log": None,
  "pc_mg_type": "full",
  "mg_levels": {
    "ksp_type": "chebyshev",
    "ksp_max_it": 2,
    "ksp_norm_type":"unpreconditioned",
    "ksp_convergence_test": "skip",
    "pc_type": "python",
    "pc_python_type": #*\color{blue}{"firedrake.PatchPC"}\label{line:pcpatch}*,
    "patch_pc_patch":  {
      "construct_type": "star",
      "construct_dim": 0
      }
  },
--  "mg_coarse_pc_type": "lu"--
}
\end{lstlisting}

%%%%%%%%%%%%%%%%%%%%%%%%%%%%%%
\section{The performance portability challenge}
%%%%%%%%%%%%%%%%%%%%%%%%%%%%%%

\begin{figure}[tb]
	\centering
	\includegraphics[width=0.4\textwidth]{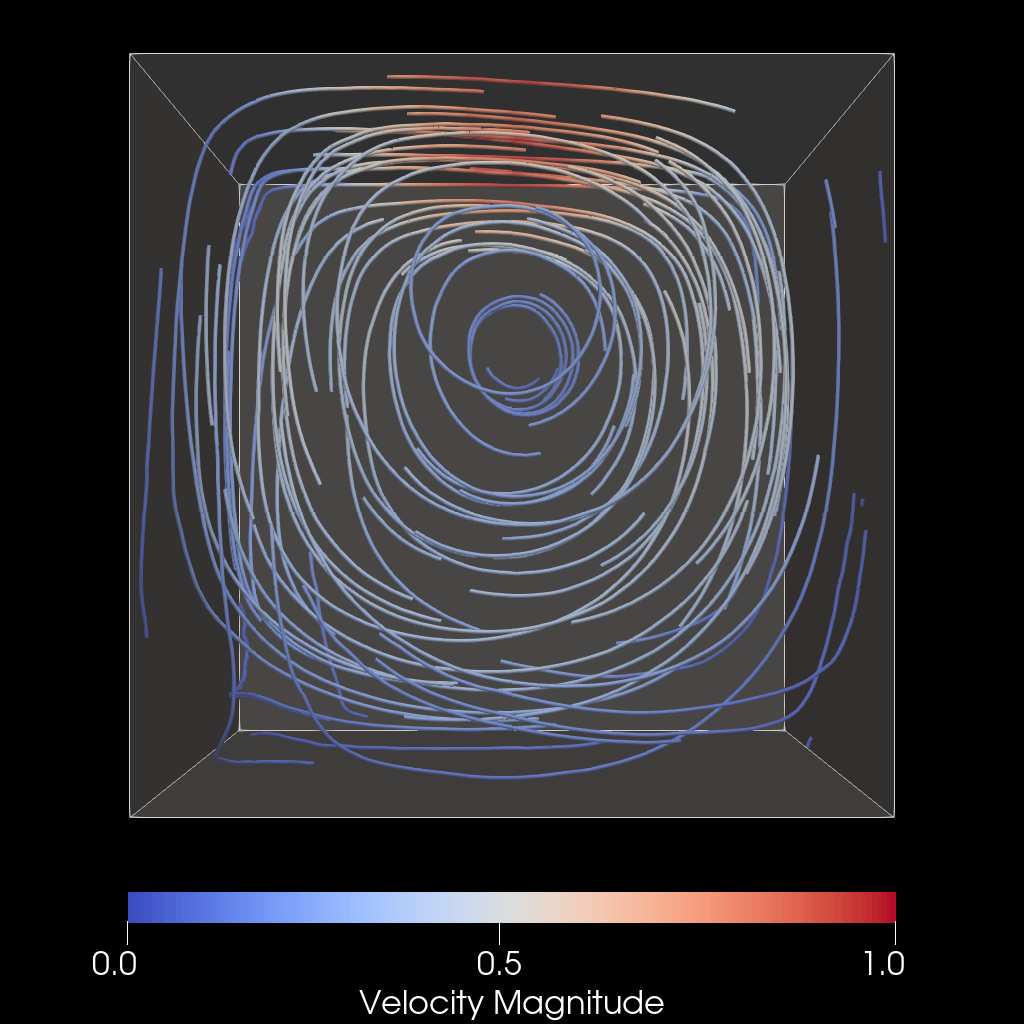}
	\vspace{-1em}
	\caption{Visualization of the velocity field in a Reynolds number 1000 lid-driven cavity Navier--Stokes simulation.}
	\label{fig:ldcpvd}
\end{figure}

Our objective will be to make effective use of different supercomputers, with important architecture differences, to produce a numerical solution to the steady incompressible Navier--Stokes equations, core equations of fluid dynamics. The scenario we will simulate is the three-dimensional lid-driven cavity (LDC) at Reynolds number 1000, a solution of which is illustrated in Figure \ref{fig:ldcpvd}. This and similar scenarios have previously been solved in Firedrake on up to one third of ARCHER, the previous UK national supercomputer \cite{farrell2020reynolds,farrell2019augmented}. However, that machine has now been decommissioned so we find ourselves in the familiar position of needing to move to a new machine.

The machines available to us are Isambard, an ARM-based supercomputer which forms part of the UK tier-2 supercomputer offering, and ARCHER2, the new UK national supercomputer which is based on AMD x86\_64 processors. At the time of writing, only a small portion of the latter machine has been commissioned. Table \ref{tab:machines} shows the characteristics of each compute node of these three machines. In comparison with ARCHER, an Isambard node has nearly three times the available memory bandwidth, and three times the core count, but each core is significantly less powerful, so there is about 50\% more bandwidth per floating point operation (FLOP). Conversely, an ARCHER2 node has twice the core count of an Isambard node, and these cores are much more powerful. The consequence is that there is about three times less bandwidth per FLOP on ARCHER2 than on Isambard. This has important consequences for how to solve our equations, as we shall see.

\begin{table}[!t]
{\centering
\begin{tabular}{l|cccc}
Machine  & CPUs & Peak DP FLOPs & Mem.BW    & Architecture\\
\hline
ARCHER   & 24  & 0.56TFLOPs & 119GiB/s & Intel x86\_64 \\
%\hline
Isambard & 64  & 1.08TFLOPs & 318GiB/s & Aarch64\\
%\hline
ARCHER2  & 128 & 4.61TFLOPs & 381GiB/s & AMD x86\_64\\
\end{tabular}
}

\caption{HPC facilities used to run these demonstrations. All values are reported \emph{per node}. ARCHER is the previous generation of HPC; Isambard has very high memory bandwidth (Mem.~BW); ARCHER2 has very high FLOP throughput.}
\label{tab:machines}
\end{table}

\section{Solving Poisson on Isambard}

Solving a steady Navier--Stokes problem at Reynolds number 1000 is a somewhat involved process. In order to ensure convergence of the nonlinear Newton iteration, a sequence of problems at increasing Reynolds number must be solved, and each of those problems is a coupled set of equations for velocity and pressure, which must be discretized using a stable pair of finite elements. The discretized nonlinear problem is solved using Newton's method, which results in a sequence of linear systems to solve. Each linear system is solved by breaking it up into separate subproblems for the velocity and pressure. In the augmented Lagrangian approach, a term is added to the equations that renders the pressure subproblem straightforward, at the cost of making the velocity solve more difficult. This velocity solve is conducted using a multigrid algorithm. Before executing this rather complicated solver, it is wise to consider the solution of the Poisson system with multigrid first. By investigating the performance of multigrid for Poisson, we can understand the consequences of the architecture in the simplest possible situation, and then use this knowledge in the full Navier--Stokes system. 

We therefore adapt the script for the Poisson system given in Listings \ref{lst:poisson} and \ref{lst:fmg} as our initial test problem.
We will want to solve on a fine mesh in order to correctly resolve all features of the solution, but this creates a very large, globally coupled system on which iterative solves will converge very slowly. Geometric multigrid algorithms overcome this by restricting the problem residual (misfit) to ever-coarser meshes, and using iterative smoothers at each level to reduce local error over ever-larger scales. 
On the coarsest mesh the problem is small enough that a direct solver can be employed (such as a Cholesky factorization of the linear system), which would be too expensive on the fine mesh.
The coarse corrections are transferred to the finer meshes, again smoothing the local error, until we reach the finest mesh. This is repeated a handful of times until the fully-resolved solution is recovered.

\begin{figure*}[p]
	\centering
	\includegraphics[width=0.8\textwidth]{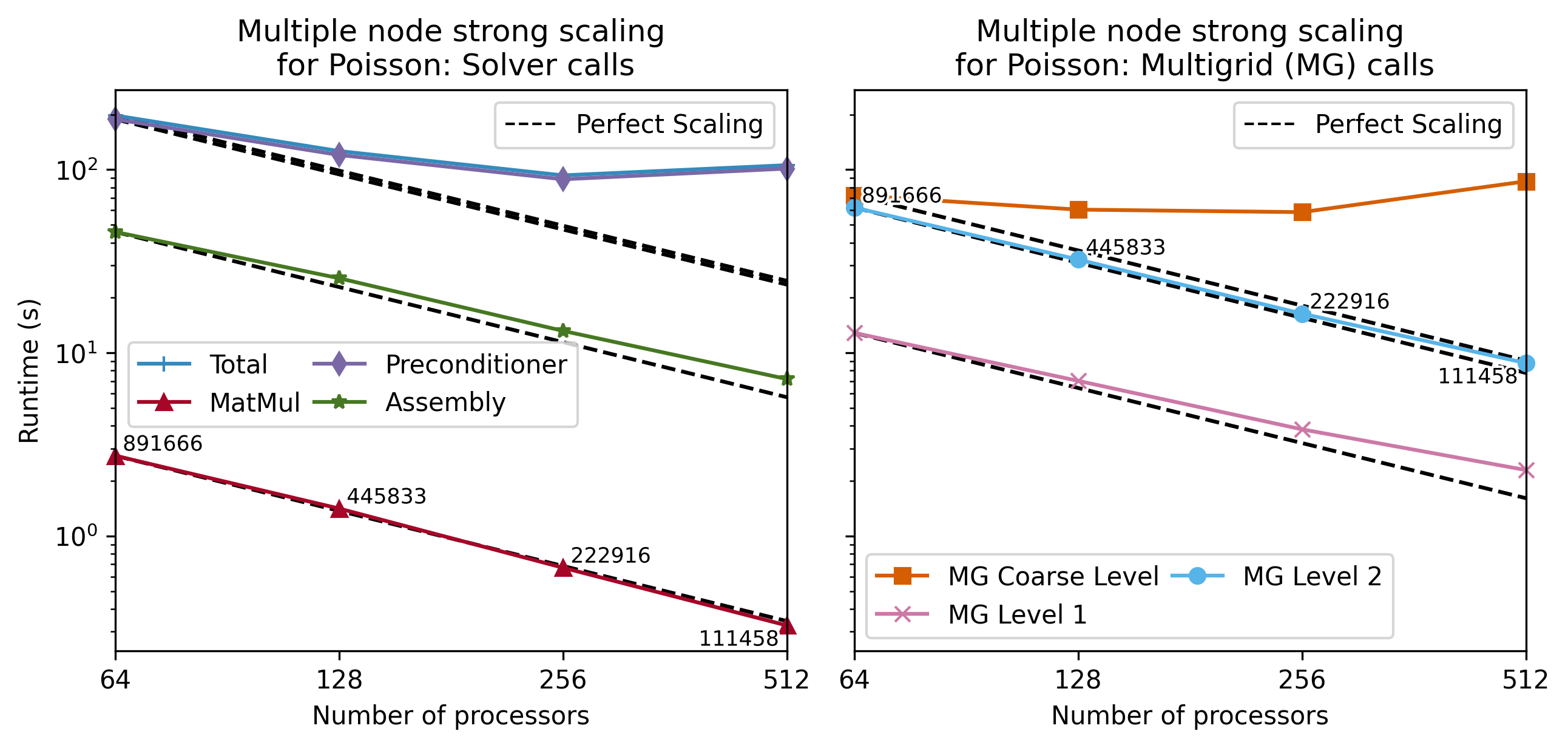}
	\vspace{-1em}
	\caption{Strong scaling of the Poisson problem on Isambard with 3 multigrid levels. The figure on the left shows the total runtime, as well as the breakdown into various component processes. The right panel shows a breakdown of the (dominant) multigrid preconditioner time into coarse solve and the two multigrid smoothing laters. The numbers on the lowest line of each graph show the DOF count per processor for the fine mesh. The coarse solver scales very poorly, and an alternative approach is required.}
	\label{fig:poisson3}
\end{figure*}

\begin{figure*}[p]
	\centering
	\includegraphics[width=0.8\textwidth]{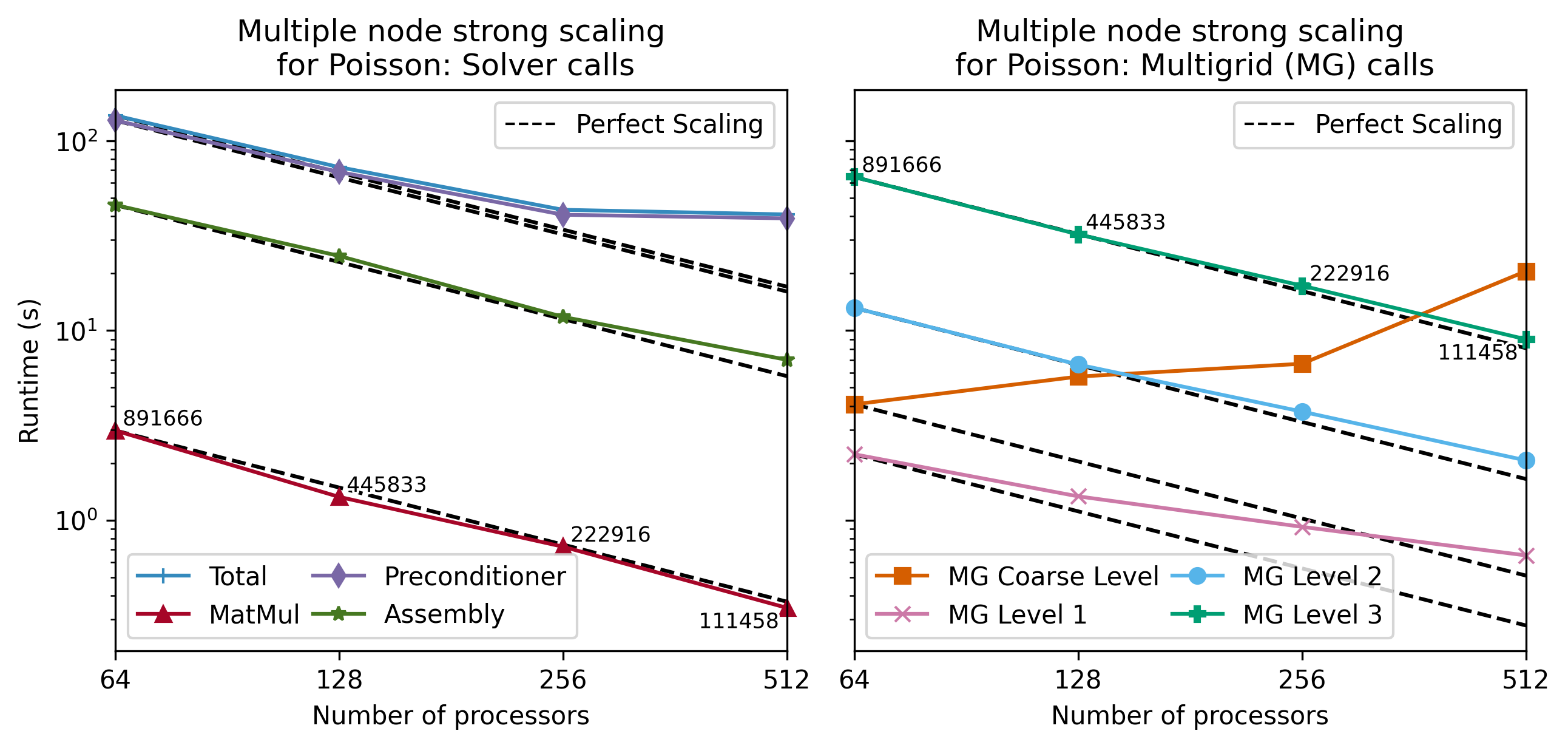}
	\vspace{-1em}
	\caption{Solving the test Poisson problem with 4 multigrid levels. The coarse solve no longer dominates on 64 cores, but it scales disastrously.}
	\label{fig:poisson4}
\end{figure*}

\begin{figure*}[p]
	\centering
	\includegraphics[width=0.8\textwidth]{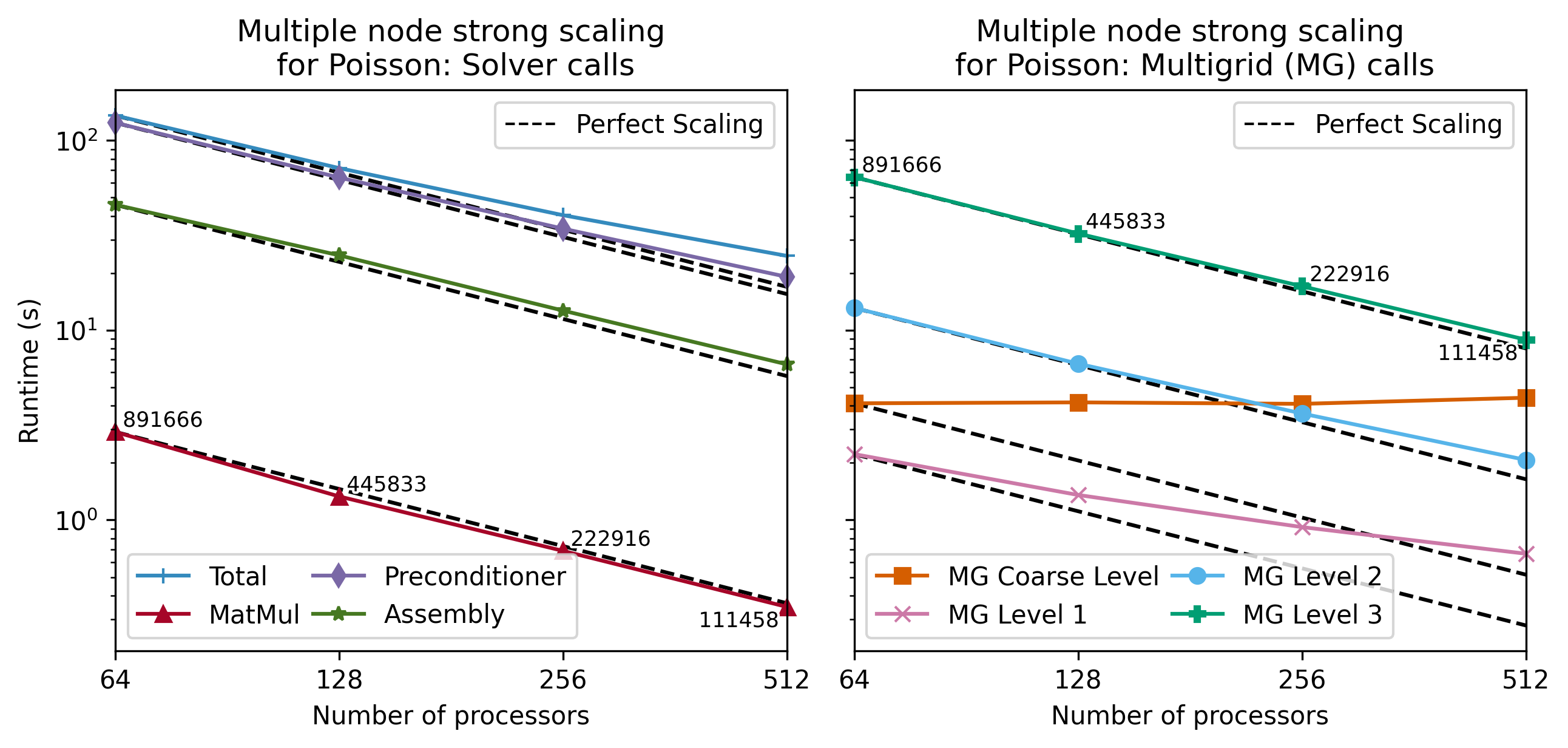}
	\vspace{-1em}
	\caption{Solving the test Poisson problem with 4 multigrid levels and telescoping the coarse grid achieves far better strong scaling.}
	\label{fig:poisson4_telescope}
\end{figure*}

%We start by examining the performance of the Poisson solver (Listing \ref{lst:poisson}, Equation (\ref{eq:poisson})).
%To get the best performance out of the multigrid solver, we must consider the hardware we are running on.
%All of the grids in the multigrid hierarchy are distributed over multiple compute nodes, and as these grids are coarsened, the number of degrees of freedom (DOFs) in the local problem gets smaller and smaller.
%Nodes in modern supercomputers have more cores than previous HPC nodes, so problems must have a large number of DOFs and high arithmetically intensity to keep all cores fully utilized.

%On the coarse grid, each core has very few degrees of freedom on the coarse grid and information must be communicated across all ranks.
%In addition, typically each node shares perhaps one or two network interface cards (NICs), which means that inter-node communication becomes a bottleneck on the coarsest mesh.
%This means, when running on Isambard, all 64 cores of each node try to communicate at once causing network contention.
%At the same time the CPU cores may not have sufficient work to keep them occupied to ameliorate the communication through the NIC.

Figure \ref{fig:poisson3} shows the time taken by the Poisson problem as it is strongly scaled on Isambard, from one to four 64-processor nodes. In a strong scaling problem, the total problem size (about 57 million DOFs in this case) is held constant while the number of processors employed is increased. Eventually, the non-parallelizable overheads in the simulation code and the similarly limited network bandwidth or latency will dominate the calculation, with the consequence that the code stops speeding up as the number of processors increases. A simple model for this effect is Amdahl's law, which models execution time as:
\begin{equation}
    t = s + p/n
\end{equation}
where $s$ is the time taken by scalar, nonparallelizable work $p$ is the time taken by parellelizable work, and $n$ is the number of processors. For an optimal multigrid solver, $p$ is expected to be linear in the total DOFs in the problem, so the appropriate unit of strong scaling is the number of DOFs per processor. This quantity will be used to measure strong scaling throughout this paper.

If we turn our attention first to the left plot of Figure \ref{fig:poisson3}, the three subcomponents of the solve shown each demonstrate their own performance story. The red line shows the amount of time taken by sparse matrix-vector multiplications in the multigrid stages in PETSc. These are, unsurprisingly, very fast and scale exceptionally well. The green line shows the cost of assembling the matrices and residuals used in the various solver stages. This is the stage of the process at which the Python layer, and the execution of code generated in C come together. However, the execution of Python code is duplicated across all processors, so were there a significant performance overhead associated with the use of Python, it would show up as a strong scaling problem here. In fact, the strong scaling of this component of the solve time is excellent. 

The same observation cannot, regrettably, be made about scaling of the multigrid solution time. This dominates the solve time, which is expected, but as configured completely fails to scale. The right plot in Figure \ref{fig:poisson3} shows the decomposition of this time into two multigrid smoothers on different meshes, and the coarse solve. The smoothers show very good scaling with the finer level 2 taking more time and the coarser level 1 scaling less well. It should be remembered that on 512 processes, multigrid level 1 is down to around 14000 DOFs per processor (=111458/8). 

The elephant in the room, however, is that the coarse solve dominates and does not scale at all. Our first response to this is to note that the idea of multigrid is to coarsen until the coarse solve is small and inexpensive, so we should coarsen further. In Firedrake this can be achieved by changing precisely two numbers in Listing \ref{lst:poisson}. On line \ref{line:meshresolution}, 32 is changed to 16 to halve the resolution of the coarse problem, and on line \ref{line:meshhierarchy}, 2 is changed to 3 to introduce an additional refinement (and hence multigrid level). Figure \ref{fig:poisson4} shows the strong scaling behaviour of the modified system. The significant change, other than the additional multigrid level, is that the cost of the coarse solve on 64 processors has dropped from about 70 seconds to about 4 seconds, which is not unexpected since the problem will have 8 times fewer degrees of freedom. However, the scaling behaviour is still awful, and by 512 processors the coarse solve takes 20 seconds and once more dominates the solve time.

This situation can be rectified using another algorithmic change. The problem here is not that our coarse solve is too large, but rather that it parallelizes poorly. This can be remedied by \emph{telescoping} the coarse solve: duplicating the problem on each node and solving them independently~\cite{may2016}. This can be achieved by changing the PETSc solver options. The red line in Listing \ref{lst:fmg} is replaced by the green lines in Listing \ref{lst:telescope}. The results can be seen in figure \ref{fig:poisson4_telescope}, which shows very good strong scaling of the whole solver out to 111458 DOFs per core.
The tight integration between PETSc and Firedrake enables sophisticated and effective solver strategies to be applied without costly recoding of solver algorithms.

\begin{lstlisting}[float={t}, caption={Telescoped full multigrid PETSc solver options. This causes the coarse problem to be solved in duplicate on each node. \lstinline+args.telescope_factor+ is a variable that is set to yield one coarse solve per node.}, label={lst:telescope}]
solver_parameters = {
  "ksp_type": "preonly",
  "pc_type": "mg",
  "pc_mg_log": None,
  "pc_mg_type": "full",
  "mg_levels": {
    "ksp_type": "chebyshev",
    "ksp_max_it": 2,
    "ksp_norm_type":"unpreconditioned",
    "ksp_convergence_test": "skip",
    "pc_type": "python",
    "pc_python_type": #*\color{blue}"firedrake.PatchPC"*,
    "patch_pc_patch":  {
      "construct_type": "star",
      "construct_dim": 0
      }
  },
+++ "mg_coarse": {
+   "pc_type": "python",
+   "pc_python_type": #*\color{blue}"firedrake.AssembledPC"*,
+   "assembled": {
+     "mat_type": "aij",
+     "pc_type": "telescope",
+     "pc_telescope_reduction_factor":
+	  	args.telescope_factor,
+     "pc_telescope_subcomm_type":
+	    "contiguous",
+	  "telescope_pc_type": "lu"
+   }
+ }++
}
\end{lstlisting}

\section{Solving Navier--Stokes on Isambard}

\begin{table*}[ht]
\centering
\begin{tabular}{c|ccccc}
\toprule
\textbf{Discretization} & \textbf{FE Pair} & \textbf{Expected} & \textbf{Stabilization} &  \textbf{Refinement} & \textbf{Smoother} \\
& velocity -- pressure & \textbf{Convergence}\\
\midrule
\parbox[c]{8em}{\centering Extended\\ Bernardi-Raugel} &
	$[\mathbb{P}_1 \oplus \mathbb{B}_3^\text{F}]^3 - \mathbb{P}_0$ &
	1--2 &
    SUPG &
	uniform &
	\parbox[c]{8em}{\centering additive star patches\\ with GMRES}\\
\midrule
\parbox[c]{8em}{\centering Scott--Vogelius} &
	$[\mathbb{P}_3]^2 - \mathbb{P}_2^\text{disc}$ &
	4 &
    interior penalty &
	barycentric &
	\parbox[c]{8em}{\centering additive \emph{macro}\\ star patches\\ with GMRES}\\
\bottomrule
\end{tabular}

\caption{The two stable finite element (FE) pairings (for the velocity and pressure fields) for the Navier--Stokes equations used in our simulations. The Scott--Vogelius pair gives faster convergence at the cost of more DOFs per element and a more expensive multigrid smoother. Expected convergence refers to the order of convergence of the velocity in the $L^2(\Omega)$ norm.}
\label{tab:ns_fem}
\end{table*}

Having fixed the scaling of the Poisson solver, we now turn our attention to the Navier--Stokes equations. We follow the Reynolds-robust solver approach given in \cite{farrell2020reynolds} and \cite{farrell2019augmented}. We therefore only sketch the solver algorithm here; however, the full code executed to conduct the experiments in this paper can be found in \cite{betteridge2021}.

Table \ref{tab:ns_fem} summarizes the two discretizations we consider for Navier--Stokes.
Initially we solve using a low-order extended Bernardi-Raugel (EBR) pairing ($[\mathbb{P}_1  \oplus \mathbb{B}_3^F]^3$--$\mathbb{P}_0$), which is a cheap and stable pairing with few degrees of freedom (DOFs) per cell.
An alternative is to use the Scott--Vogelius (SV) pairing $[\mathbb{P}_3]^2$--$\mathbb{P}_2^\text{disc}$. This converges at higher order, and exactly
enforces the incompressiblity constraint, which is crucial at high Reynolds numbers~\cite{john2017}.
Table \ref{tab:ns_fem} also highlights that using an alternative discretization requires changing many other components,
including the stabilization used, the mesh hierarchy employed, and the multigrid relaxation on each level.

\begin{figure*}[htp]
	\centering
	\includegraphics[width=\textwidth]{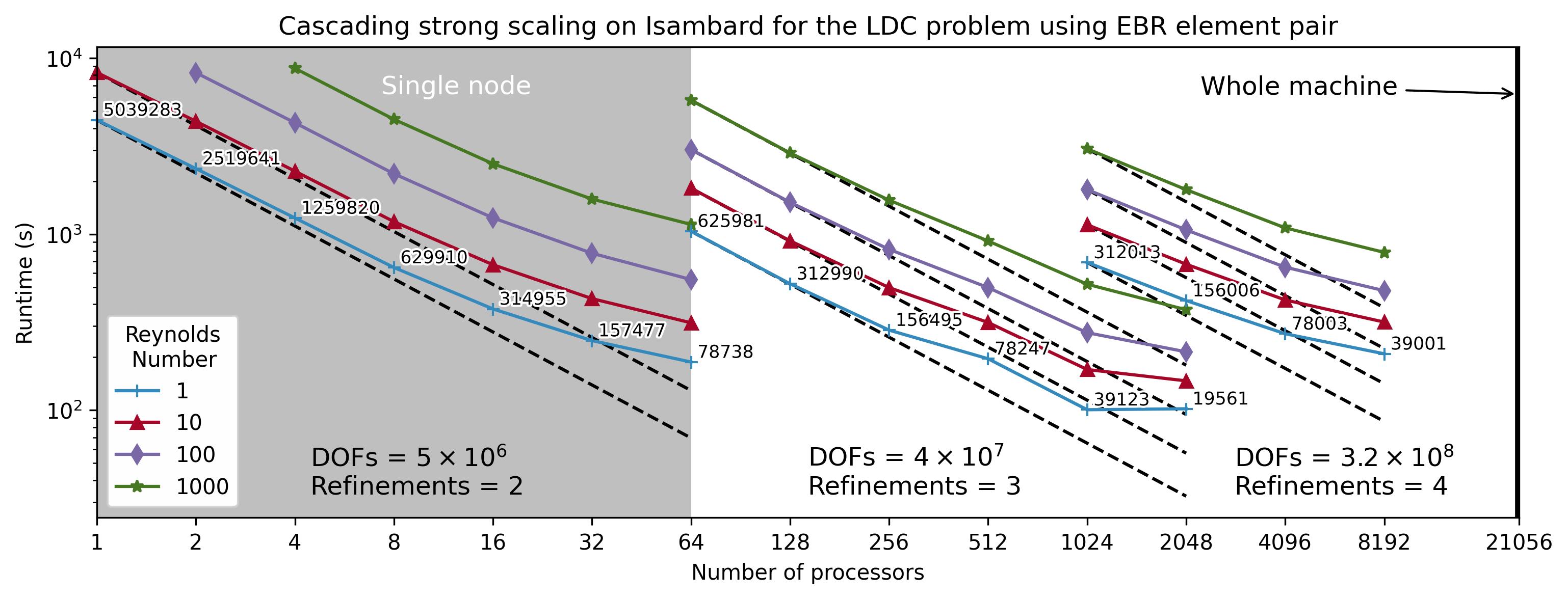}
	\caption{A sequence of strong scaling results for the lid-driven cavity on Isambard. In each case, reasonably good strong scaling is observed down to tens of thousands of DOFs per processor. The absence of ideal scaling lines for the highest Reynolds numbers on a single node is caused by the absence of a single processor run for these cases: it is difficult to justify occupying a whole node for many hours to solve a problem in serial that would take minutes to run in parallel.}
	\label{fig:cascading}
\end{figure*}

With the knowledge gained from solving Poisson we were able to adjust the PETSc solver options for the Navier--Stokes LDC application and successfully utilize more than a third of Isambard.
Strong scaling results for three different problem sizes (5, 40 and 320 million DOFs) with the EBR discretization are shown in Figure \ref{fig:cascading}.
We solve the problem up to Reynolds number 1000 using numerical continuation; the scaling for intermediate solutions are also plotted, but the top green line with star markers corresponds to the most numerically challenging solution to obtain. 
The figure shows consistent strong scaling up to somewhere between $40\ 000$ and $80\ 000$ DOFs per core on all three problem sizes.

%%%%%%%%%%%%%%%%%%%%%%%%%%%%%%
\subsection{Solving Navier-Stokes on ARCHER2}
\begin{figure*}[htp]
	\centering
	\includegraphics[width=\textwidth]{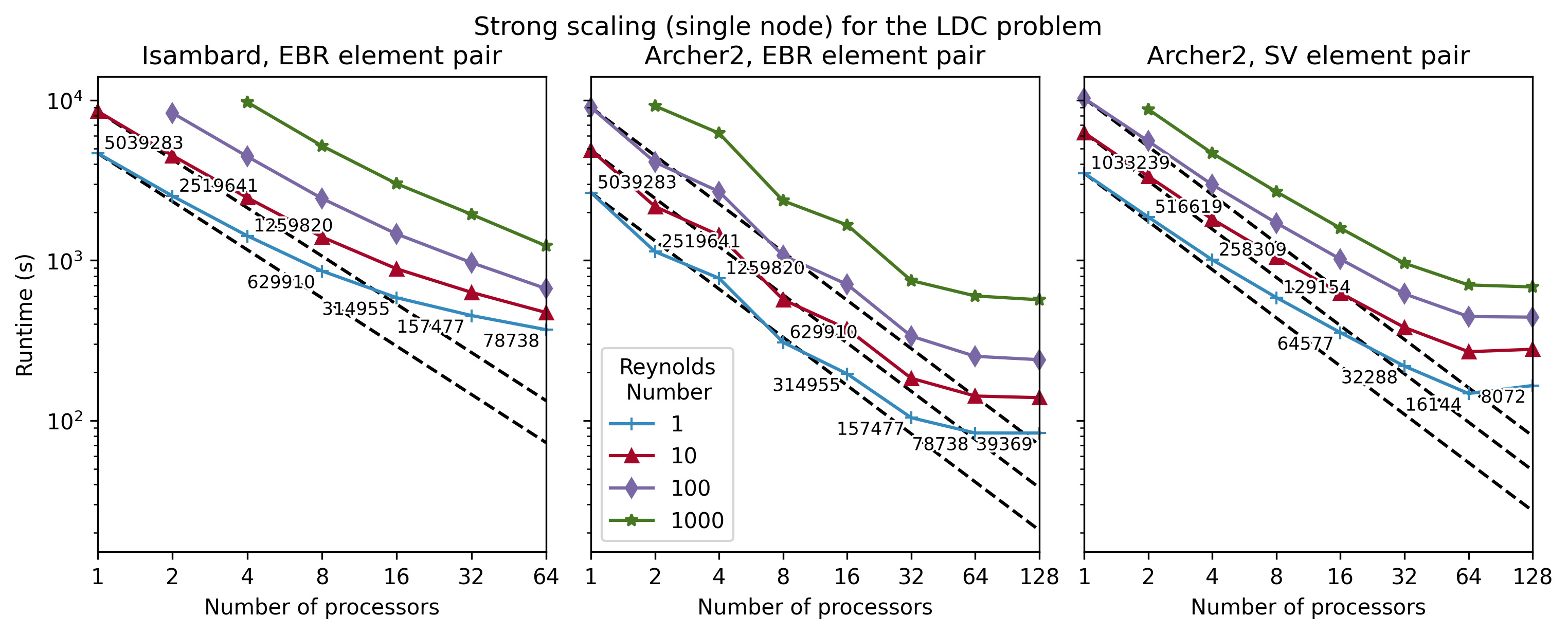}
	\caption{Single node strong scaling results: Isambard low order (EBR, left), ARCHER2 low order (EBR, middle), and ARCHER2 high order (SV, bottom) discretizations}
	\label{fig:ldc_sys_comp}
\end{figure*}

%%%%%%%%%%%%%%%%%%%%%%%%%%%%%%
At the time of writing the ARCHER2 HPC facility is not fully available and the existing portion of the machine is still in the evaluation phase.
As a result, we are only currently able to report single node performance on this machine.
This still allows us to perform an ``apples to apples'' comparison of Isambard and ARCHER2 single node performance, which is shown for the LDC Navier--Stokes problem in Figure \ref{fig:ldc_sys_comp}.

In the left and middle plot, identical problems (size, discretization, and solver options) are solved on Isambard and ARCHER2 respectively, and we see two distinct behaviors.
On Isambard, in the left plot, we observe good strong scaling performance. In particular the top green line with star markers (corresponding to the highest Reynolds number) scales well. At lower Reynolds number the scaling is not as good, but this is a less challenging computation and hence a less interesting case.
The J-shaped runtime curve is indicative of Amdahl's law in action: as the number of processors increases the non-paralellizable workload smoothly takes an increasing proportion of runtime.
On ARCHER2, in the middle graph, there is an abrupt halt to scaling and a distinct ``hockey stick'' shape.
Beyond approximately 32 cores the simulation stops strong scaling, this suggest that we have run out of another resource; in this case, there is not enough memory bandwidth to continue strong scaling. This means that our seemingly quite good strong scaling curves are hiding our failure to effectively utilize the abundance of FLOPs on an ARCHER2 node.

When a finite element simulation is limited by memory bandwidth, one way to improve the strong scaling is to increase the degree of the finite element used in the discretization, which increases the arithmetic intensity of the simulation, pushing us back towards being compute bound.
For the Poisson benchmark problem it is straightforward to change polynomial degree, for example by changing line \ref{line:degree} of Listing \ref{lst:poisson} from 3 to 4 (or higher).
For the Navier--Stokes equations, the situation is different. The elements used for velocity and pressure must satisfy a compatibility condition for stability and cannot be chosen independently.
To obtain a more accurate solution we change from the EBR finite element pair to the Scott--Vogelius pair, and make the necessary modifications to the stabilization, mesh hierarchy, and multigrid relaxation, as outlined in Table \ref{tab:ns_fem}.

\begin{figure}[htp]
	\centering
	\includegraphics[width=0.5\textwidth]{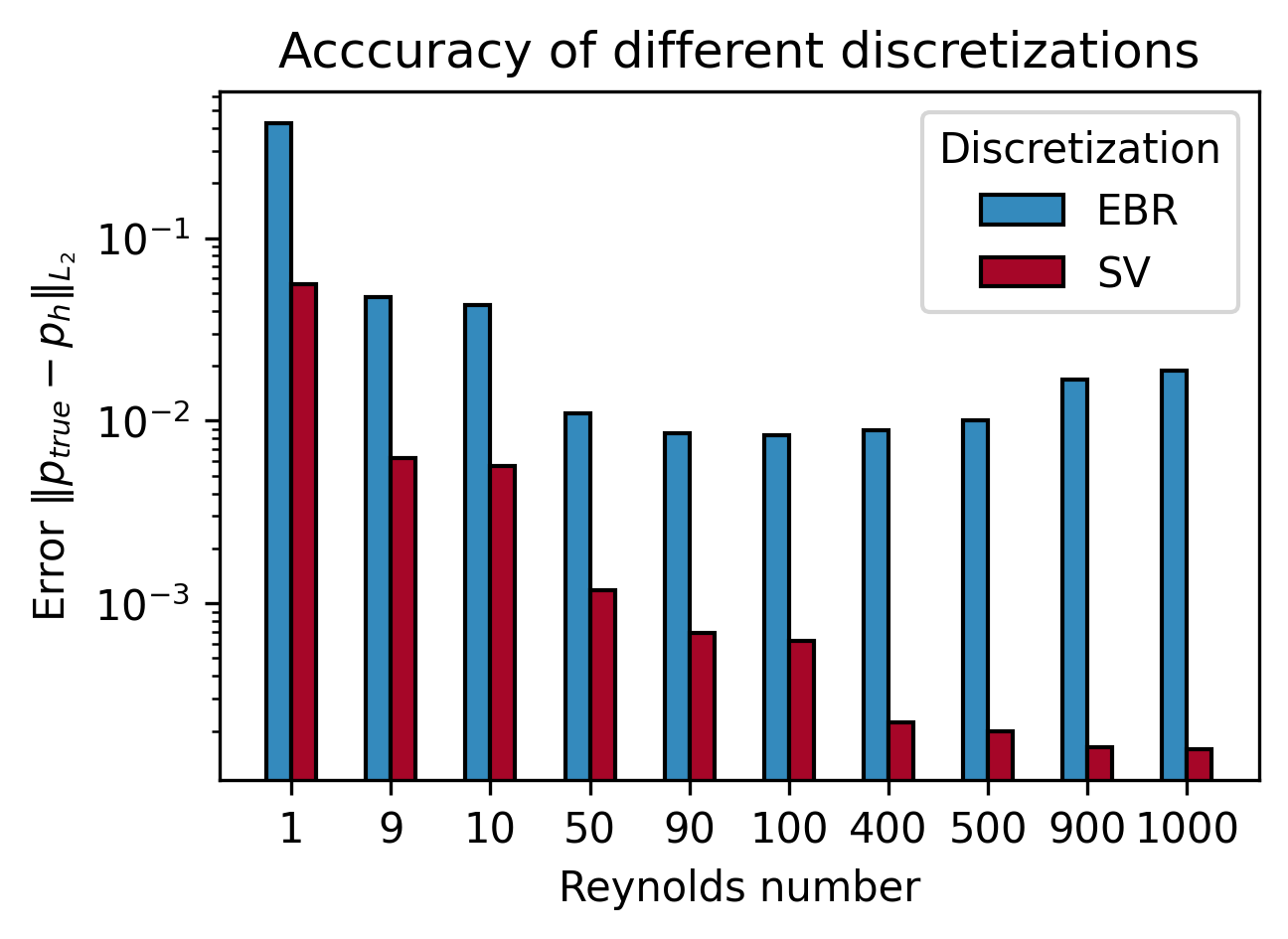}
	\caption{Norm of the difference between the computed solution and the analytic solution for the pressure field at the different Reynolds numbers used in the continuation steps for the LDC problem.}
	\label{fig:error}
\end{figure}

Figure \ref{fig:ldc_sys_comp} shows the different scaling behaviour for the Scott--Vogelius ($[\mathbb{P}_3]^3$--$\mathbb{P}_2^\text{disc}$) discretization in the right graph.
The problem size has to be reduced  to fit on a single node. One might initially observe that, whilst the strong scaling has improved, the total number of DOFs for the problem has significantly decreased (5 million down to 1 million), but the runtime has not changed.
However, Figure \ref{fig:error} demonstrates the utility of this new discretization: the more expensive per DOF algorithm obtains a significantly more accurate solution.

%%%%%%%%%%%%%%%%%%%%%%%%%%%%%%
\section{CONCLUSION}
%%%%%%%%%%%%%%%%%%%%%%%%%%%%%%

Making effective use of whichever supercomputing resources are available to a scientist or engineer at any given time demands that simulation algorithms, and not just their implementations, are adapted to the available hardware. The changes presented here in porting to Isambard and ARCHER2 span the discretization used, choice and configuration of preconditioner, stabilization scheme, and refinement algorithm. These changes were required for just one application in fluid dynamics to be effectively ported between three machines with relatively conventional CPU-based architectures. The immense diversity of continuum mechanics simulation challenges across science and engineering and their ever-changing nature means that the space of algorithmic combinations that will be required by different users is essentially unbounded. Further, the amount of developer time available to achieve acceptable performance for a given application is typically severely limited.

Here we have demonstrated that, using Firedrake, it is possible to achieve this combination of short, expressive code that can be easily modified, and scalable parallel performance. The key to this is the tight integration of code generation for the discretized operators with the composable, programmable solvers and preconditioners provided by PETSc. Using Python as the user-level language facilitates the expressivity of the user code, while code generation and PETSc's callback mechanisms ensure that the expensive inner loops are optimized, compiled code. A further advantage of this composable approach is that developments created for one application become immediately available to other applications without the need for costly re-implementation.

Firedrake has already been applied in many applications resulting in hundreds of papers, but there are always limits to its capabilities. Among the many directions in which it could be extended, support for adaptive mesh refinement, local polynomial order refinement, and support for GPUs are developments currently under consideration that would be particularly useful in high performance computing contexts.

%%%%%%%%%%%%%%%%%%%%%%%%%%%%%%
\section{ACKNOWLEDGMENTS}
%%%%%%%%%%%%%%%%%%%%%%%%%%%%%%
This work used the Isambard UK National Tier-2 HPC Service (\url{http://gw4.ac.uk/isambard/}) operated by GW4 and the UK Met Office, and funded by the Engineering and Physical Sciences Research Council (EPSRC) (EP/P020224/1). This work used the ARCHER2 UK National Supercomputing Service (\url{https://www.archer2.ac.uk}).
This work was supported by the United Kingdom Research and Innovation (UKRI) ExCALIBUR program (EPSRC grant EP/V001493/1).

\nocite{*}
%\begin{thebibliography}{1}
\bibliographystyle{IEEEtran}
\bibliography{references}

%\end{thebibliography}
\newpage

\begin{IEEEbiography}{Jack D.~Betteridge,}{\,} is a research software engineer at Imperial College London.
His research interests lie in code generation applications for numerically solving PDEs on HPC.
His doctorate is from the University of Bath.
Contact him at j.betteridge@imperial.ac.uk.
\end{IEEEbiography}

\begin{IEEEbiography}{Patrick E.~Farrell,}{\,}is an associate professor in the numerical analysis group at the University of Oxford. His research interests are in fast preconditioners for PDEs, and in bifurcation analysis of nonlinear problems. He was jointly awarded the 2015 Wilkinson Prize for Numerical Software for the development of the dolfin-adjoint automated inverse simulation system, a 2015 Leslie Fox prize in numerical analysis, and the 2021 Charles Broyden prize in optimization. He holds a doctorate from Imperial College London. Contact him at patrick.farrell@maths.ox.ac.uk.
\end{IEEEbiography}

\begin{IEEEbiography}{David A.~Ham,}{\,} is a reader in computational mathematics at Imperial College London. His research interests are in the development of high level abstractions and code generation technology for PDEs. He was jointly awarded the 2015 Wilkinson Prize for Numerical Software. He is chief-executive editor of the journal Geoscientific Model Development. He holds a doctorate from Delft University of Technology.  Contact him at david.ham@imperial.ac.uk.
\end{IEEEbiography}

\vfill{}

\end{document}